\input{aipcheck}

\documentclass[
  ,draft            
  ]
  {aipproc}

\layoutstyle{6x9}
\usepackage{array}
\usepackage{amsmath}
\usepackage{amssymb}
\def\vspa{\vspace{-1cm}}
\def\vspb{\vspace{-.5cm}}
\def\vspc{\vspace{-.25cm}}
\def\vspf{\vspace{-7cm}}

\begin{document}

\title[DE-DM Interaction and violation of EP.]{Dark Energy-Dark Matter Interaction from the Abell Cluster A586 and
violation of the Equivalence Principle}

\classification{98.80.-k,98.80.Cq,95.35.+d,95.36.+x,98.65.Dx,04.20.Cv}
\keywords      {Cosmology -- dark matter -- dark energy -- Abell Cluster
A586 -- Equivalence principle}

\author{Morgan Le Delliou}{
  address={ {\bf Speaker}; C
F
T
C
, U
. Lisboa 
 Av. Gama Pinto 2, 1649-003 Lisboa, Portugal}
}

\author{Orfeu Bertolami}{
  address={Centro de F\'{i}sica dos Plasmas, IST, Lisbon, Portugal}
}

\author{Francisco Gil Pedro}{
  address={Departamento de F\'{\i}sica, Instituto Superior T\'{e}cnico, Lisbon, Portugal}
}

\begin{abstract}
We find that the Abell Cluster A586 exhibits evidence of the interaction
between dark matter and dark energy and argue that this interaction
suggests evidence of violation of the Equivalence Principle. This violation is
found in the context of two different models of dark energy-dark matter
interaction.
\end{abstract}

\maketitle

~

\vspa\vspb
\section{Introduction\vspc}

Contemporary cosmology regards consensually the nature of dark energy
and dark matter (hereafter DE and DM) to be crucial for a suitable
description of universe's evolution and dynamics.
Even though observations agree with the $\Lambda$CDM
model, a deeper insight into DE and DM leads to
more complex models -- in particular, interaction between them.
However, so far, no evidence of this interaction has been presented.
In this work, we argue that Cluster A586 exhibits such
evidence. Furthermore, we argue that this suggests evidence of violation
of the Equivalence Principle (EP). 

In what follows, we first set the formalism for DE-DM interaction
and consider two very different, phenomenologically viable models:
one based on \textit{ad hoc} DE-DM interaction \cite{Amendola},
the other on the generalized Chaplygin gas (GCG) model with explicit
identification of DE and DM \cite{Bento04}. Our observational
inferences are based on Cluster A586, given its stationarity,
spherical symmetry and wealth of available observations \cite{Cypriano:2005}.
We also compare our results with other cosmological observations \cite{Guo07}.\vspb

\section{Interacting models\vspc}

Our results are obtained in the context of two distinct phenomenologically
viable models for the DE-DM interaction: a naturally unified model, the GCG \cite{Bento02}, but also a less
constrained interacting model with constant DE equation
of state (hereafter EOS) parameter $\omega_{DE}=p_{DE}/\rho_{DE}$
(where $p_{X}$ is pressure and $\rho_{X}$ is energy density
of $X$; see e.g. \cite{Amendola}).

We consider first a quintessence model with constant EOS. The Bianchi
conservation equations for both DE and DM read ($H$ denoting the
Hubble parameter and $\zeta$, the interaction strength)\vspa
\begin{center}\begin{eqnarray}
\dot{\rho}_{DM}+3H\rho_{DM} & = & \zeta H\rho_{DM},\\
\dot{\rho}_{DE}+3H\rho_{DE}(1+\omega_{DE}) & = & -\zeta H\rho_{DM},\end{eqnarray}
\par\end{center}
where constant EOS parameter ($\omega_{DE}$), and scaling ($\eta$)
given by
$\frac{\rho_{DE}}{\rho_{DM}}=\frac{\Omega_{DE_{0}}}{\Omega_{DM_{0}}}a^{\eta}$
are assumed,
as in \cite{Amendola} (with density parameter $\Omega_{X}$ -- $\rho_{X}$
scaled with critical density -- and $a$, the scale of the universe).
Then, the coupling varies as
$\zeta=-\frac{(\eta+3\omega_{DE})\Omega_{DE_{0}}}{\Omega_{DE_{0}}+\Omega_{DM_{0}}a^{-\eta}}$
and 
energy densities evolve as
\vspa\vspc
\begin{center}\begin{align}
\rho_{{\scriptscriptstyle DM}}= & a^{-3}\rho_{{\scriptscriptstyle DM_{0}}}\left[\Omega_{{\scriptscriptstyle DE_{0}}}a^{\eta}+\Omega_{{\scriptscriptstyle DM_{0}}}\right]^{-\frac{(\eta+3\omega_{{\scriptscriptstyle DE}})}{\eta}}, & \rho_{{\scriptscriptstyle DE}}= & a^{\eta}\rho_{{\scriptscriptstyle DE_{0}}}\frac{\rho_{{\scriptscriptstyle DM}}}{\rho_{{\scriptscriptstyle DM_{0}}}}.\end{align}
\par\end{center}\vspc
We turn now to the GCG model. It is defined by its unified EOS $p_{DE}=-A/(\rho_{DM}+\rho_{DE})^{\alpha}$,
with $A$ and $\alpha$ constant, and assuming DE constant EOS
parameter $\omega_{DE}=-1$ \cite{Bento02}. This last splitting into DE and
DM, together with Bianchi conservation imply a scaling behaviour with 
$\eta=3(1+\alpha)$ and energy densities to be\vspa
\begin{center}\begin{align}
\rho_{DM} & =\rho_{DM_{0}}\frac{\rho_{DE}}{\rho_{DE_{0}}}a^{-3(1+\alpha)}, & \rho_{DE}= & \rho_{DE_{0}}\left(\frac{\Omega_{DE_{0}}+\Omega_{DM_{0}}}{\Omega_{DE_{0}}+\Omega_{DM_{0}}a^{-3(1+\alpha)}}\right)^{\frac{\alpha}{1+\alpha}}.\end{align}
\par\end{center}

\section{Generalized Layzer-Irvine equations\vspb}

In this work we shall focus on the effect of interaction on clustering
as revealed by the Layzer-Irvine equation. We write the kinetic and
potential energy densities $\rho_{K}$ and $\rho_{W}$ of clustering DM in interacting models
in terms of scale factor dependence\vspb\vspc
\noindent \begin{center}\begin{align}
\rho_{K} & \propto a^{-2}, & \rho_{W} & \propto a^{\zeta-1},\end{align}
\par\end{center}\vspc
and we use DM virialization dynamics in an expanding
universe described by the generalised Layzer-Irvine equation
\cite{Peebles}\vspb\vspc
\noindent \begin{center}\begin{align}
\dot{\rho}_{DM}+\left(2\rho_{K}+\rho_{W}\right)H & =\zeta H\rho_{W},\end{align}
\par\end{center}\vspc
with $\zeta$, the coupling for the quintessence model (for the GCG, 
the scaling allows to just replace $\eta=3(1+\alpha)$ and
$\omega_{DE}=-1$). 
At equilibrium, $2\rho_{K}+\rho_{W}=\zeta\rho_{W}\ne0$.\vspb

\section{Detection of interaction from observation of A586\vspb}

Using those models, we have interpreted observations from an Abell
cluster \cite{Cypriano:2005} and compared it with other observations
\cite{Guo07}.

%
\begin{figure}
  \begin{centering}
  \includegraphics[height=.3\textheight]{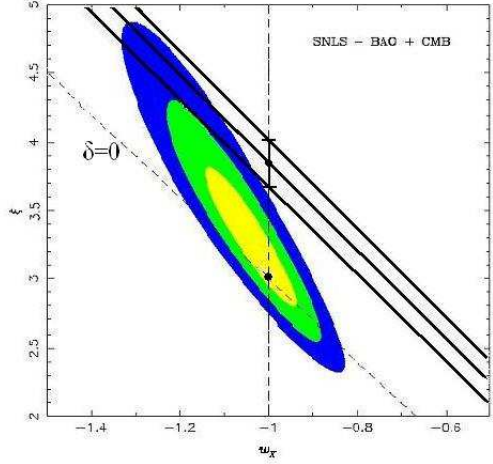}
  \end{centering}
  \caption{\label{fig:GuoPlus}Superimposition of the probability contours for
the interacting DE-DM model in the ($\omega_{X}$,$\xi$) plane (denoted
as ($\omega_{DE_{0}}$,$\eta$) in \cite{Berto07}), marginalized
over $\Omega_{DE_{0}}$ in \cite{Guo07} study of CMB, SN-Ia and BAO
(2.66<$\xi$<4.05 at 95\% C.L.) with the results extended from \cite{Berto07},
based on the study of A586 cluster. The $\xi=-3\omega_{X}$ line corresponds
to uncoupled models.}
\end{figure}
From \cite{Cypriano:2005},
we extract: the total mass, $M_{Cluster}=(4.3\pm0.7)\times10^{14}\: M_{\odot}$
(galaxies, DM and intra-cluster gas), radius, $R_{Cluster}=$ 422 kpc
at $z=0.1708$ (angular radius $\Delta_{max}=145''$) and cluster
dispersion $\sigma_{v}=(1243\pm58)\: kms^{-1}$, from
weak lensing. From photometry, we obtain $<R>=\frac{2}{N_{gal}(N_{gal}-1)}\sum_{i=2}^{N_{gal}}{\sum_{j=1}^{i-1}r_{ij}}$,
the mean intergalactic distance%
\footnote{where $r_{ij}^{2}=2d^{2}\left[1-\cos(\alpha_{ci}-\alpha_{cj})\cos\delta_{ci}\cos\delta_{cj}-\sin\delta_{ci}\sin\delta_{cj}\right]$,
$\alpha_{ci}=\alpha_{i}-\alpha_{center}$ and $\delta_{ci}=\delta_{i}-\delta_{center}$%
}, from declinations and right assentions of a galaxy sub-sample
within the projected $\Delta_{max}$ (i.e. for galaxy $i$, $\sqrt{\alpha_{ci}^{2}+\delta_{ci}^{2}}\leq\Delta_{max}$).\enlargethispage{.25cm}
Further assuming $\omega_{DE}=-1$ and 
$\Omega_{DE_{0}}=0.72$, $\Omega_{DM_{0}}=0.24$
\cite{Spergel:2006hy}, we get \vspc\begin{align}
\rho_{K} & \simeq\frac{9}{8\pi}\frac{M_{Cluster}}{R_{Cluster}^{3}}\sigma_{v}^{2}=(2.14\pm0.55)\times10^{-10}Jm^{-3},\\
\rho_{W} & \simeq-\frac{3}{8\pi}\frac{G}{<R>}\frac{M_{Cluster}^{2}}{R_{Cluster}^{3}}=(-2.83\pm0.92)\times10^{-10}Jm^{-3},\end{align}
which allow us to obtain \cite{Berto07}\vspc\begin{align}
\eta & =3.82_{-0.17}^{+0.18}\neq-3\omega_{DE}, & \alpha & =0.27_{-0.06}^{+0.06}\neq0.\end{align}\vspb

Notice that $\eta\neq-3\omega_{DE}$ signals the energy exchange between
DM and DE. It is also remarkable that $\alpha\neq0$ which implies
the GCG description is not degenerate with $\Lambda$CDM ($\alpha=0$).
We mention that our results (see Fig.\ref{fig:GuoPlus} and \cite{OBPedro07b})
are consistent with the study of \cite{Guo07} where DE-DM interacting
quintessence are analysed for compatibility with WMAP CMB \cite{Spergel:2006hy},
SNLS SN-Ia \cite{SNLS06} and Baryon Acoustic Oscillations in SDSS
\cite{BAOinSDSS}.\vspb

\section{Violation of Equivalence Principle?\vspc}

Given that the EP concerns the way matter falls in the gravitational
field, considering the clustering of matter against the cosmic expansion
and the interaction with DE seems to be a logical way to test its
validity. Both models \emph{predict} departure of homogeneous DM from
dust behaviour and have effects that can be interpreted as violation
of EP.%
\begin{figure}
\begin{centering}\begin{tabular}{r>{\raggedright}m{0.35\paperwidth}}
(a)\includegraphics[height=0.3\textheight]{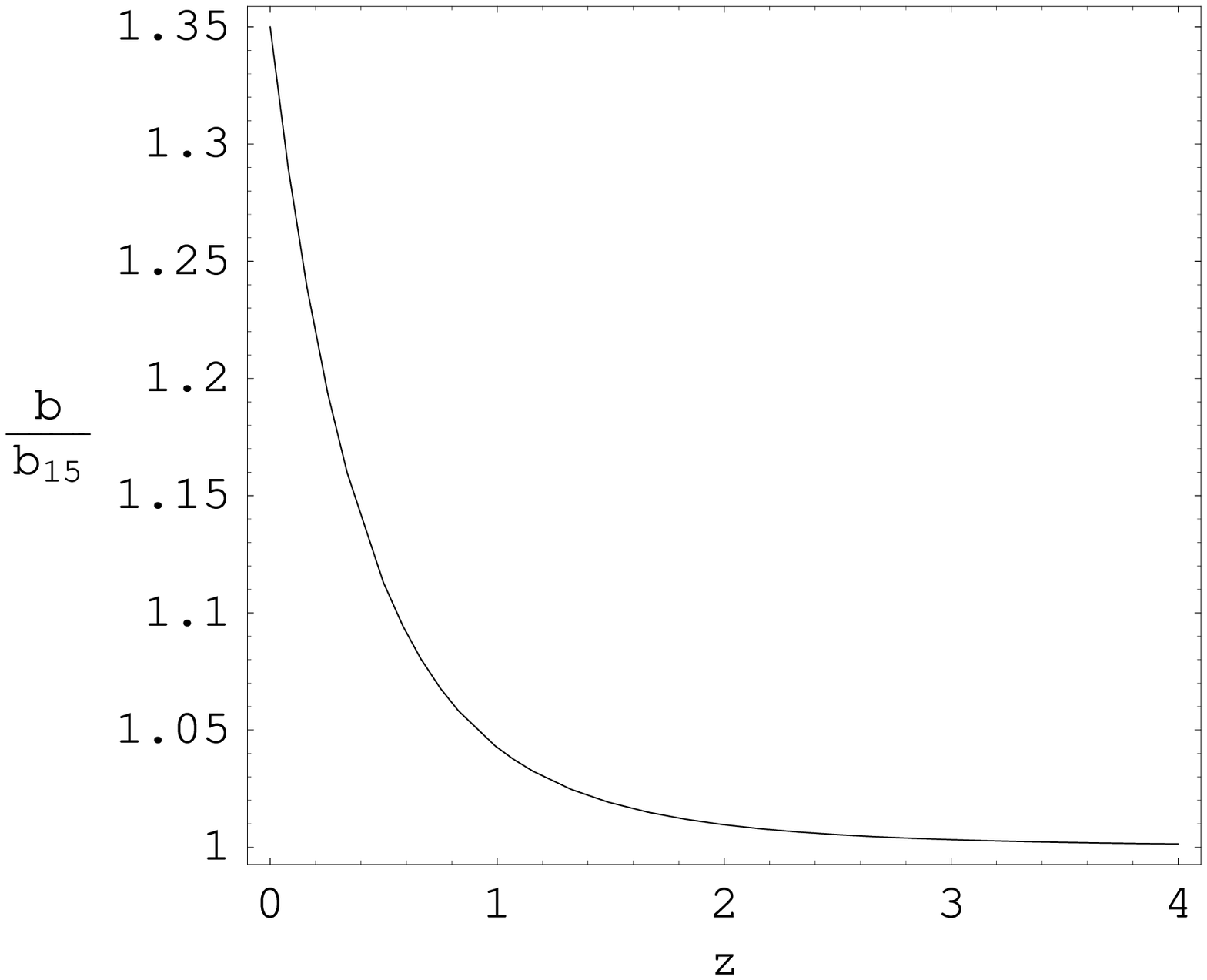}&
\parbox[t]{0.35\paperwidth}{~\vspf\\
\includegraphics[height=0.33\textheight]{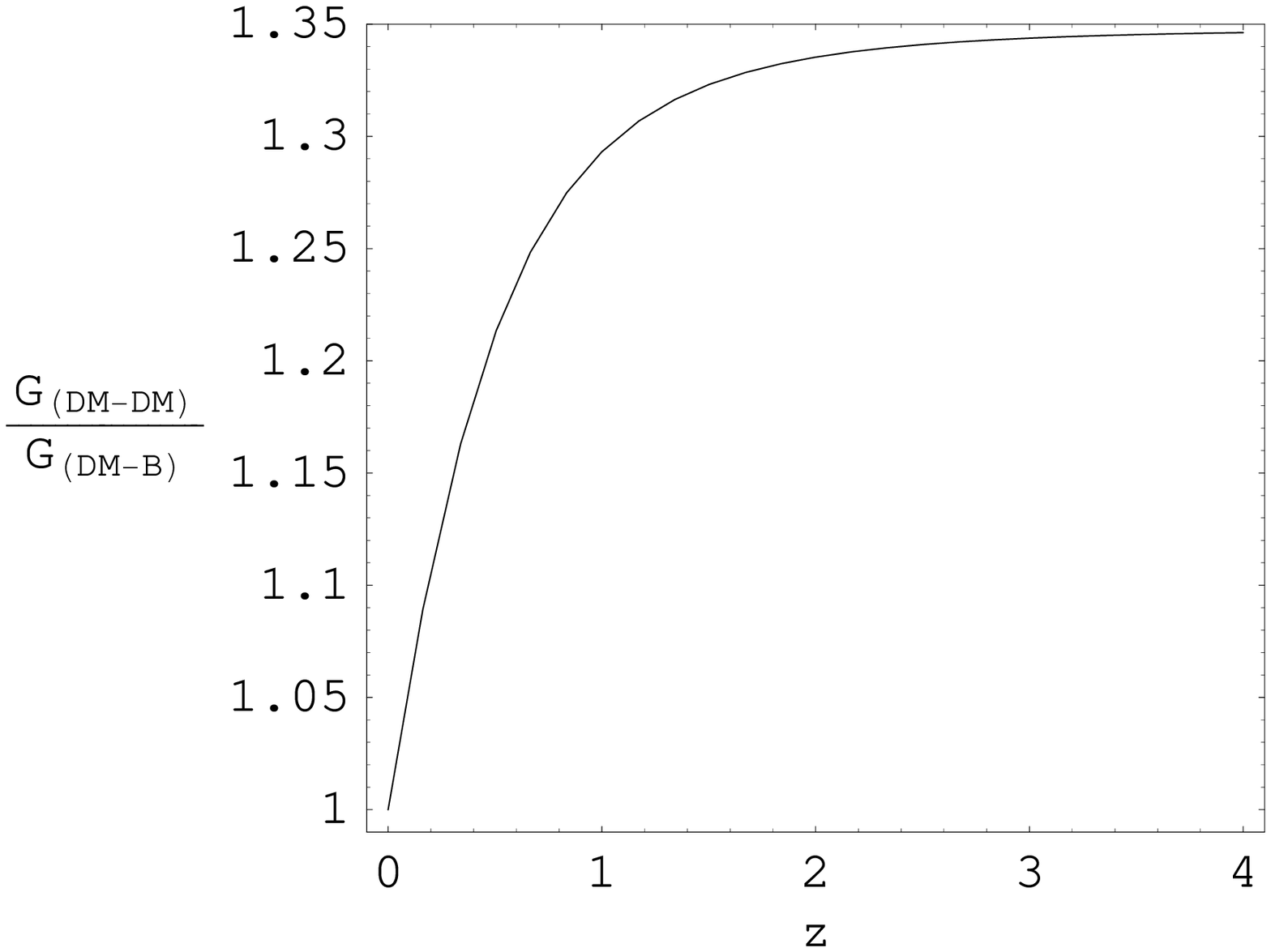}\vspb\vspc\\
(b)}\tabularnewline
\end{tabular}\par\end{centering}
\caption{\label{fig:BiasParam}(a) Normalized gravitationally induced bias parameter
as a function of $z$, where $b_{15}\equiv b(z=15)$, $z=15$
is the typical condensation scale and $b={\rho_{B}/\rho_{DM}}=\Omega_{B_{0}}/\Omega_{DM_{0}}[\Omega_{DE_{0}}a^{\eta}+\Omega_{DM_{0}}]^{(\eta+3\omega_{DE})/\eta}.$
\label{fig:diffG}(b) Evolution with redshift of the ratio of the
gravitational coupling for DM and baryons falling on a DM halo, using
the varying coupling model discussed in \cite{OBPedro07b}, to be
compared with the simulation of \cite{KesdenKamion06}. $G_{(DM-DM)}=G_{(DM-B)}(\Omega_{DE_{0}}a^{\eta}+\Omega_{DM_{0}})^{-(\eta+3\omega_{DE_{0}})/\eta}.$}
\end{figure}

The non-dust evolution of DM density leads to evolution of the bias
parameter, $b=\rho_{B}/\rho_{DM}$, at the homogeneous level on cosmological
timescales \cite{Berto07} (Fig. \ref{fig:BiasParam}a). Other astrophysical
effects also affect the bias so the detection of this drift would
require statistics over different $z$ ranges.

\noindent If we attribute the non-dust $\rho_{DM}$ to dust-like DM
particles mass,\vspc\begin{align}
m_{DM}(a)= & m_{DM,0}[\Omega_{DE_{0}}a^{\eta}+\Omega_{DM_{0}}]^{-\frac{\eta+3\omega_{DE_{0}}}{\eta}},\end{align}
gravity is then Baryon/DM composition dependent as seen in the two
particles potential $U_{(1-2)}=-\frac{Gm_{1}m_{2}}{r_{12}}$. We can
now assign the time evolution to a varying G, as seen in Fig. (\ref{fig:diffG}b)%
 \cite{OBPedro07b}, to compare with simulations of the type done
by \cite{KesdenKamion06}.\vspb
\newpage
~
\vspa\vspa\vspc
\section{Conclusions\vspb}

Observations of Cluster A586 \cite{Cypriano:2005} suggest
evidence of departure from virialization given that A586 is very
spherical and relaxed (from its mass distribution and Gyrs
without mergers). The generalized Layzer-Irvine equation allows
to interpret this departure as interacting DE. We therefore link the
observed virialization to interaction \cite{Berto07,OBPedro07b} with
two different models, consistent with known
constraints \cite[refs. therein]{Berto07}: an interacting
quintessence with constant $\omega_{DE}$ {\tiny }\cite{Amendola}
and a Chaplygin gas with $\omega_{DE}=-1$ {\tiny }\cite{Bento02}.

From these models, we argue that the Equivalence Principle should
be violated as they impact, for example, on the overall bias parameter
\cite{Berto07}
and on Baryon/DM asymmetric collapse \cite[using][]{KesdenKamion06,OBPedro07b}.
This is consistent with violation that other models have reported
\cite{Alimi06}.\vspb\vspb\enlargethispage{.5cm}


\begin{theacknowledgments}
\vspb
The work of MLeD is supported by FCT (Portugal),
SFRH/BD/16630/2004 and hosted by J.P. Mimoso and CFTC, Lisbon University
.\vspb\vspc
\end{theacknowledgments}





\IfFileExists{\jobname.bbl}{}
 {\typeout{}
  \typeout{******************************************}
  \typeout{** Please run "bibtex \jobname" to optain}
  \typeout{** the bibliography and then re-run LaTeX}
  \typeout{** twice to fix the references!}
  \typeout{******************************************}
  \typeout{}
 }


%
%
%
%
%
%

\end{document}